\documentstyle[12pt]{article}

\begin{document}

\centerline{\bf\Large SO(10)-Inspired See-Saw Mechanism}

\vspace{2 cm}

\centerline{\large{{Mario Abud$^*$ and Franco Buccella$^{\circ}$}}}

\vspace{1 cm} \centerline{$^*$ Dipartimento di Scienze Fisiche
dell'Universit\`a di Napoli}\ \centerline{Complesso Universitario di M. St.
Angelo}\ \centerline{via Cintia, I-80126 Napoli} \ \centerline{and INFN, Sezione
di Napoli}\ \centerline{\bf e-mail :Abud@na.infn.it}\newline

\centerline{$^{\circ}$ Theory Division, CERN, CH-1211 Geneva 23, Switzerland
and}\ \centerline{\ Istituto di Fisica Teorica, Universit\`{a} di Napoli,Italy}\
\centerline{{\bf e-mail :Buccella@na.infn.it}}

\vspace{2 cm}
\begin{abstract}
\noindent

We determine the $\nu_{R}$ Majorana mass matrix from the experimental data
on neutrino oscillations in the framework of a see-saw SO(10) model, where
we impose the condition $(M^R)_{33}$ = 0 to avoid too large fine-tunings in
the see-saw formula. We find a class of solutions with the two lowest
neutrino masses almost degenerate and the scale of the matrix elements of $%
M^R$ in the range $10^{11} -10^{12}$ GeV in agreement with Pati-Salam
intermediate symmetry. We find also solutions with smaller neutrino masses,
for which the scale of $M_R$ depends on the solution to the "solar neutrino
problem" and on the value of the component of $\nu_e$ along the highest mass
eigenstate, $U_{e3}$.
\end{abstract}

\textwidth=20cm \textheight=22.5cm

\renewcommand{\theequation}{\arabic{section}.\arabic{equation}} \topmargin %
-1.5cm \oddsidemargin -0.2cm \evensidemargin -0.2cm





\vspace{1 cm}

\newpage

\noindent 

\setcounter{page}{1} \pagestyle{plain}

\section{Introduction}

The evidence in favour of oscillations\cite{P} of solar\cite{sun} and
atmospheric \cite{atm} neutrinos with square mass differences orders of
magnitude smaller than the square masses of the other fermions strongly
supports SO(10) gauge unification\cite{so10}, where the see-saw mechanism%
\cite{ss} naturally accounts for this large mass gap. In this framework it
is important to establish the order of magnitude of the elements of the $\nu
_{R}$'s Majorana mass matrix,which is related to the scale of the
spontaneous symmetry breaking of the SO(10) generator, B - L \cite{CMP}.%
\newline
In the following section we shall show the uncertainties for the $\nu _{L}$
effective Majorana mass matrix,which follow not only from the presence of
different solutions to the ''solar neutrino problem''\cite{BKS}, but also
from not knowing the mass of the lightest neutrino mass eigenstate and the
relative sign of the three mass eigenvalues. We shall also show that an
expansion in the eigenvalues of the Dirac neutrino mass matrix, assumed to be
hierarchically ordered according to the family as the other fermions, is
unable to reproduce with the first two terms the phenomenological properties
of the $\nu _{L}$ effective Majorana mass-matrix, if small mixing angles for
the Dirac mass matrices are assumed for leptons in analogy with what we know
about the CKM matrix\cite{CKM} for quarks.\newline
This fact motivates us to take in the third section a vanishing value for $%
M_{33}^{R}$, \cite{js}\cite{St}\cite{ABFRT}. Within the simplifying assumption
of a trivial mixing matrix for the Dirac leptons, we are able to implement
that condition for all the different solutions of the ''solar neutrino
problem''. In most cases we find solutions with the two lower masses almost
degenerate and at the order of magnitude of the highest one and with the
highest matrix element of $M^{R}$,$M_{23}^{R}$, in the range $%
10^{11}-10^{12} $ GeV in good agreement with the scale of the spontaneous
breaking of B-L,~ $2.7\cdot 10^{11}$ GeV \cite{BP} in the model with
SU(4)xSU(2)xSU(2) \cite{ps} intermediate symmetry.We find also solutions
with the two lower masses at the order~$\sqrt{\Delta m_{sun}^{2}}$ and in
that case the order of magnitude of the largest matrix element of $M^{R}$
depends on the solution to the ''solar neutrino problem'' and on the value
of $U_{e3}$.\newline
Finally we give our conclusions.

\section{ The effective $\nu_L$ Majorana mass matrix.}

\setcounter{equation}{0}

The experimental information from solar and atmospheric neutrinos is such to
"almost" allow a model-independent determination of the neutrino mass
matrix, within the hypothesis of the existence of only the three active
Majorana neutrinos and CP symmetry. The second hypothesis is expected to be
at most an approximation, since CP violation in the neutral kaon system most
probably comes from a phase in the CKM matrix, as is confirmed by the
indication of a large CP-violating asymmetry in the $J/\psi \, K^S$ \cite
{CDF} channel, and a similar phase may be present in the lepton sector, for
which, with Majorana neutrinos, two other CP-violating phases are also
allowed.

The incompleteness in the determination of the neutrino mass matrix arises
from the following facts:

\begin{itemize}
\item[i)]  Up to now there is only an upper limit for $|U_{e3}|^{2}(\le .05)$
\cite{ch}\cite{bg1}.

\item[ii)]  Oscillations depend on $\Delta m^{2}$'s and cannot by themselves
determine the masses of the neutrino eigenstates.

\item[iii)]  There are sign ambiguities in the mass-eigenvalues.
\end{itemize}

Moreover there are five solutions \cite{BKS} of the ``neutrino solar
problem" with the oscillation parameters reported in Table I.

\begin{center}
\begin{tabular}{|c|c|c|c|}
\hline
&  &  &  \\ 
Scenario & $\frac{\Delta m_{s}^{2}}{eV^{2}}$ & $sin^{2}(2\theta _{s})$ & C.L.
\\ \hline
&  &  &  \\ 
LMA & 2.7 $\cdot 10^{-5}$ & 0.79 & 68 \% \\ \hline
&  &  &  \\ 
SMA & 5.0 $\cdot 10^{-6}$ & 0.72 $\cdot 10^{-2}$ & 64 \% \\ \hline
LOW & 1.0 $\cdot 10^{-7}$ & 0.91 & 83\% \\ \hline
$VAC_{S}$ & 6.5 $\cdot 10^{-11}$ & 0.72 & 90 \% \\ \hline
$VAC_{L}$ & 4.4 $\cdot 10^{-10}$ & 0.90 & 95 \% \\ \hline
\end{tabular}
\bigskip \\[0pt]
Table I
\end{center}

For atmospheric neutrino oscillations the relevant parameters are \cite{atm}%
: 
\begin{equation}
\Delta m^2_a \simeq 3.5\cdot10^{-3} (eV)^2 \quad \quad \quad
\sin^2(2\theta_a) \simeq 1  \label{1}
\end{equation}
We shall use this information to determine the effective $\nu_L$ Majorana
mass matrix.

By assuming CP symmetry, which implies a real symmetric $3\times 3$ Majorana
neutrino mass matrix $M^{L}$ : 
\[
M^{L}=\left( m_{ij}\right) _{i,j=1,2,3} 
\]
\begin{equation}
=\Sigma \;|m_{i}>m_{i}<m_{i}|  \label{2}
\end{equation}
where the eigenvalues $m_{i}$ are real, not necessarily positive, numbers.

Our convention is 
\begin{equation}
|m_{3}|>|m_{2}|>|m_{1}|  \label{3}
\end{equation}
and we may take $m_{3}>0$, since an overall change of sign of $M^{L}$ has no
physical consequences.

We define 
\begin{eqnarray}
|m_3>&=& \sin \psi |\nu_e> + \cos \psi (\cos \theta |\nu_\mu> + \sin \theta
|\nu_\tau >)  \nonumber \\
|m_2>&=& - \sin \chi |v_1> + \cos \chi |v_2>  \nonumber \\
|m_1>&=& \cos \chi |v_1> + \sin \chi |v_2>  \label{4}
\end{eqnarray}
where 
\begin{eqnarray}
|v_1>&=& \cos \psi |\nu_e> - \sin \psi (\cos \theta |\nu_\mu> + \sin \theta
|\nu_\tau >)  \nonumber \\
|v_2>&=& - \sin \theta |\nu_\mu> + \cos \theta |\nu_\tau >  \label{5}
\end{eqnarray}
The difference $m^2_3 - m^2_1$ has to be identified with $\Delta m^2_a$,
while for $\Delta m^2_s$ one has two options, either $m_3^2 - m_2^2$ or $%
m_2^2 - m_1^2$. The current interpretation of the atmospheric neutrino
anomaly due to a practically maximal mixing between $\nu_\mu$ and $\nu_\tau$%
, with $\nu_e$ playing a marginal role, would imply with the first choice
having $\nu_e$ components almost completely along the two nearly degenerate
higher-mass eigenstates. Since this situation is unnatural in the $SO(10)$
framework, which we will describe in the second part of this paper,we take
the option $\Delta m^2_s = m_2^2 -m^2_1$.

From eqs. \ref{2}, \ref{4} and \ref{5}, it is easy to get: 
\begin{eqnarray}
m_{11} &=&m_{3}\sin ^{2}\psi +m_{2}\sin ^{2}\chi \cos ^{2}\psi +m_{1}\cos
^{2}\chi \cos ^{2}\psi  \nonumber \\
m_{22} &=&m_{3}\cos ^{2}\theta \cos ^{2}\psi +\cos ^{2}\theta \sin ^{2}\psi
(m_{2}\sin ^{2}\chi +m_{1}\cos ^{2}\chi )  \nonumber \\
&+&sin^{2}\theta (m_{2}\cos ^{2}\chi +m_{1}\sin ^{2}\chi )-\sin 2\theta \sin
2\chi \frac{\sin \psi }{2}(m_{2}-m_{1})  \nonumber \\
m_{33} &=&m_{3}\sin ^{2}\theta \cos ^{2}\psi +sin^{2}\theta \sin ^{2}\psi 
(m_{2}\sin ^{2}\chi +m_{1}\cos ^{2}\chi )  \\ \label{6}
&+&\cos ^{2}\theta (m_{2}\cos ^{2}\chi +m_{1}\sin ^{2}\chi )+\sin 2\theta
\sin 2\chi \frac{\sin \psi }{2}(m_{2}-m_{1})  \nonumber \\
m_{12} &=&\cos \psi \biggl[ \sin \psi \cos \theta (m_{3}-m_{2}\sin ^{2}\chi
-m_{1}\cos ^{2}\chi ) \nonumber \\
&~& \qquad\qquad + \sin \theta \frac{\sin 2\chi }{2}(m_{2}-m_{1})
\biggr]  \nonumber \\
m_{13} &=&\cos \psi \biggl[\sin \psi \sin \theta (m_{3}-m_{2}\sin
^{2}\chi-m_{1}\cos ^{2}\chi ) \nonumber \\
&~& \qquad\qquad -\cos \theta \;\frac{\sin \,2\chi }{2}%
\;(m_{2}-m_{1}) \biggr]  \nonumber \\
m_{23} &=&\frac{\sin \;2\theta }{2}\;\biggl[m_{3}\cos ^{2}\psi -m_{2}\cos ^{2}\chi
-m_{1}\sin ^{2}\chi  \nonumber \\
& & \qquad\qquad +\sin ^{2}\psi (m_{2}\sin ^{2}\chi +m_{1}\cos ^{2}\chi )\biggr]  \nonumber \\
&+& \cos 2\theta \;\frac{\sin 2\chi }{2}\;\sin \psi  (m_{2}-m_{1}) \nonumber 
\end{eqnarray}The angles $\theta $ and $\chi $ should be identified with the mixing angles
for atmospheric and solar neutrino oscillations, respectively. The
experimental value found for the atmospheric neutrino oscillations, $\sin
^{2}2\theta _{a}\simeq 1$, and the upper limit on $|U_{e3}|^{2}$ imply $%
\theta =\frac{\pi }{4}$ and a small value for $\sin \psi $.

With hierarchical neutrino masses 
\begin{equation}
m_3 >> |m_2| >> |m_1|  \label{7}
\end{equation}
one has 
\begin{eqnarray}
m^2_3 & \simeq & \Delta m^2_a  \nonumber \\
m^2_2 & \simeq & \Delta m^2_s  \label{8}
\end{eqnarray}
and by taking $\psi= 0$ for the contributions proportional to $m_2$ and
neglecting $|m_1 | << \sqrt{\Delta m^2_s}$, one gets:

\begin{eqnarray}
M^{L} &=&\frac{m_{3}}{2}\,\cos ^{2}\psi \left( 
\begin{array}{ccc}
2\tan \psi ^{2} & \sqrt{2}\tan \psi & \sqrt{2}\tan \psi \\ 
\sqrt{2}\tan \psi & 1 & 1 \\ 
\sqrt{2}\tan \psi & 1 & 1,
\end{array}
\right)  \nonumber \\
&+&\frac{m_{2}}{2}\cos ^{2}\chi \left( 
\begin{array}{ccc}
2\tan ^{2}\chi & \sqrt{2}\tan \chi & -\sqrt{2}\tan \chi \\ 
\sqrt{2}\tan \chi & 1 & -1 \\ 
-\sqrt{2}\tan \chi & -1 & 1,
\end{array}
\right) .  \label{9}
\end{eqnarray}
We now consider the possibility, often advocated, that the two lower masses
are almost degenerate and larger than $\sqrt{\Delta m_{s}^{2}}$, but smaller
than $\sqrt{\Delta m_{a}^{2}}$ \cite{js} \cite{St} \cite{ABFRT}. In that
case, one has still $\Delta m_{a}^{2}=m_{3}^{2}$, but we can neglect $\Delta
m_{sun}^{2}$ with respect to $m_{1}^{2}$ and $m_{2}^{2}$. The first term in
eq. \ref{9} remains the same, while the other one takes a different form,
depending on the relative sign of $m_{2}$ and $m_{1}$.We will consider $%
m_{1}m_{2}<0$, since this assumption will be necessary in the following and
get 
\begin{equation}
\frac{m_{2}\cos 2\chi }{2}\,\left( 
\begin{array}{ccc}
1 & \sqrt{2}\tan 2\chi & -\sqrt{2}\tan 2\chi \\ 
\sqrt{2}\tan 2\chi & 1 & -1 \\ 
-\sqrt{2}\tan \ 2\chi & -1 & 1,
\end{array}
\right)  \label{10}
\end{equation}
Again for the term proportional to $m_{2}$ in eq. \ref
{10}  we made the approximation to take $\psi =0$.

We now consider the possibility advocated by Georgi and Glashow \cite{GG},
with the motivation of having a sizeable neutrino contribution to the hot
dark matter, of almost degenerate square masses for the neutrinos, larger
than their differences. We have four options, according to the relative
signs of the $m_{i}$'s. By taking again $m_{1}<0<m_{2}$ and $\psi =0$ for the
part proportional to $\Delta m_{a}^{2}$, we get : 
\begin{eqnarray}
M^{L} &=&m_{3}\cos ^{2}\chi \left( 
\begin{array}{ccc}
\tan ^{2}\chi -1 & \sqrt{2}\cos \psi \tan \chi & -\sqrt{2}\cos \psi \tan \chi
\\ 
\sqrt{2}\cos \psi \tan \chi & 1 & \tan ^{2}\chi \\ 
-\sqrt{2}\cos \psi \tan \chi & \tan ^{2}\chi & 1,
\end{array}
\right)  \nonumber \\
&+&m_{3}\cos ^{2}\chi \sin \psi \left( 
\begin{array}{ccc}
0 & \sqrt{2}\cos \psi & \sqrt{2}\cos \psi \\ 
\sqrt{2}\cos \psi & -2\tan \chi & 0 \\ 
\sqrt{2}\cos \psi & 0 & 2\tan \chi ,
\end{array}
\right)  \nonumber \\
&+&2m_{3}\cos ^{2}\chi sin^{2}\psi \left( 
\begin{array}{ccc}
1 & 0 & 0 \\ 
0 & -\frac{1}{2} & -\frac{1}{2} \\ 
0 & -\frac{1}{2} & -\frac{1}{2},
\end{array}
\right)  \nonumber \\
&+&\frac{\Delta m_{a}^{2}\cos 2\chi }{4m_{3}}\left( 
\begin{array}{ccc}
2 & -\sqrt{2}\tan 2\chi & -\sqrt{2}\tan 2\chi \\ 
-\sqrt{2}\tan 2\chi & 1 & 1 \\ 
-\sqrt{2}\tan 2\chi & 1 & 1,
\end{array}
\right) .  \label{11}
\end{eqnarray}
Let us consider 
\begin{eqnarray}
m_{\nu _{e}\nu _{e}} &=&m_{3}(-\cos 2\chi +2\cos ^{2}\chi \sin ^{2}\psi 
\nonumber \\
&+&\sin ^{2}\chi \,\frac{\Delta m_{s}^{2}}{2m_{3}^{2}}+\frac{\Delta m_{a}^{2}%
}{2m_{3}^{2}}\cos ^{2}\chi )  \label{12}
\end{eqnarray}
for which there is an experimental upper limit coming from the study of
neutrinoless double beta-decay $|m_{\nu _{e}\nu _{e}}|<.2$ $eV$ \cite{beta}.
To get the cancellation between $-\cos 2\chi $ and $2\cos ^{2}\chi \sin
2\psi $, $\chi $ should be near to $\frac{\pi }{4}$, more precisely : 
\begin{equation}
\sin ^{2}\psi =\frac{\tan ^{2}\chi -1}{2}  \label{13}
\end{equation}
But the r.h.s. of eq. \ref{13} takes at least the central value .22 (for $%
\sin ^{2}2\theta _{s}=.91$), larger than the upper limit for $\sin ^{2}\psi $%
, which implies \cite{VISS} 
\begin{equation}
|m_{3}|\simeq |\frac{m_{\nu _{e}\nu _{e}}}{2\cos ^{2}\chi \sin ^{2}\psi
-\cos 2\chi }|\le .85eV  \label{14}
\end{equation}
The situation is similar for the case with $m_{1}>0>m_{2}$. In conclusion
many options are open for the neutrino mass matrix and even an exact
determination of the mixing matrix and of the square mass differences is
unable to resolve the ambiguity associated to the value of $|m_{1}|$ and to
the relative signs of the masses \cite{FF}.Also the same Majorana mass
matrix for the $\nu _{L}$'s may be obtained with different choices of the
Dirac lepton matrices and of $M^{R}$.

It has been suggested since a long time \cite{hrr} that the peculiar
properties of neutrino mixing, almost maximal for atmospheric, rather large
for all the solutions to the ``solar neutrino problem", except the MSW small
angle solution, should be related to the fact that, differently from the
case for the charged fermions, their masses arise from the see-saw mechanism 
\cite{ss}.The later has the indisputable merit of providing a reason for the
small value of the neutrino masses, especially in the framework of $SO(10)$
unified theories, which predict the existence of left-handed antineutrinos
with Majorana masses related to the spontaneous breaking of $B-L$ symmetry 
\cite{CMP}.

Unified $SO(10)$ theories are suitable for the study of fermion masses as
all the fermions of one family belong to a single representation, the
spinorial 16 \cite{so10}. Indeed, by classifying the Higgs doublet
responsible for the breaking of the electroweak symmetry in the vector
representation (10), one obtains,together with the celebrated( but actually
not so successfull) relationship \cite{SU}, 
\begin{equation}
\frac{m_{\tau}}{m_b} (unification)= 1  \label{15}
\end{equation}
while the analogous relationship, $\frac{m_{\nu_{\tau}}}{m_t} (unification)=
1$, is turned by the see-saw mechanism into the intriguing prediction of
very small neutrino masses. To correct $\frac{m_\mu}{m_s}= \frac{m_e}{m_d}
=1 $, some component of the electroweak Higgs, at least along the 126,
should be introduced \cite{GJ} 
\begin{eqnarray}
(16 \times 16)_S & =& 126 + 10  \nonumber \\
(16 \times 16)_A &=& 120  \label{16}
\end{eqnarray}
Without adopting a particular scheme, we shall limit ourselves to assume
that the Dirac neutrino mass matrix, once diagonalized by the biunitary
transformation, gives rise to a hierarchical relationship for the elements
of the diagonal matrix similar to that existing for the other fermions, and
the matrix corresponding to the CKM, which gives the misalignment with the
corresponding leptons, has small non diagonal matrix elements. 
\begin{eqnarray}
M^L &=& -m^D (M^R)^{-1} (m^D)^T  \nonumber \\
m^D &=& U_L m_{diag} U_R^+   \label{17} 
\end{eqnarray}
with 
\begin{equation}
m_{diag}= diag ( \mu_1 , \mu_2, \mu_3)  \label{18}
\end{equation}
and we assume $\mu_1 << \mu_2 << \mu_3 $. From eqs. \ref{17} it is easy to
derive: 
\begin{equation}
M^L= - U_L \, m_{diag} \, U_R^ + \, (M^R)^{-1} \, U_R^\star \, m_{diag} \,
U_L^T  \label{19}
\end{equation}
which is simplified by defining 
\begin{equation}
\tilde{N} \, = \, U^+_R\, M^{R^{-1}}\, U_R^\star  \label{20}
\end{equation}
into 
\begin{equation}
M^L= -U_L\, m_{diag}\, \tilde{N}\, m_{diag}\, U_L^T  \label{21}
\end{equation}
With our hypothesis of CP conservation, $U_R$ and $U_L$ are, in fact, real
orthogonal matrices and $\tilde{N}$ is symmetric real. We may develope $M^L$
as a quadratic form in the $\mu_i$'s: 
\begin{eqnarray}
- M^L_{\ell \ell^{\prime}} &=& (U_L)_{\ell 3} (U_L)_{\ell ^{\prime}3} \tilde{%
N}_{33} \mu^2_3  \nonumber \\
&+& ((U_L)_{\ell 2} (U_L)_{\ell ^{\prime}3} + (U_L)_{\ell 3} (U_L)_{\ell
^{\prime}2}) \tilde{N}_{23} \mu_2 \mu_3  \nonumber \\
&+& (U_L)_{\ell 2} (U_L)_{\ell ^{\prime}2} \tilde{N}_{22} \mu^2_2 + \cdots
\label{22}
\end{eqnarray}

Should the matrix elements of $\tilde N$ be of the same order, it would be a
reasonable approximation to take only the terms proportional to $\mu_3^2$,
which would give for $-M^L$ the expression: 
\begin{equation}
\tilde N_{33} \mu^2_3 \left ( \begin{matrix} {(U_L)^2_{13} & (U_L)_{13}
(U_L)_{23}& (U_L)_{13} (U_L)_{33} &\cr (U_L)_{13}
(U_L)_{23}&(U_L)_{23}^2&(U_L)_{23} (U_L)_{33} \cr (U_L)_{13} (U_L)_{33}&
(U_L)_{23} (U_L)_{33}&(U_L)_{33}^2} \end{matrix} \right )  \label{23}
\end{equation}
The matrix defined by eq. \ref{23} should have eigenvalues $\mu^2_3 \, \, 
\tilde{N}_{33}$, corresponding to the eigenvector 
\begin{equation}
\left( 
\begin{array}{c}
U_{13} \\ 
U_{23} \\ 
U_{33}
\end{array}
\right ),  \label{24}
\end{equation}
and twice the eigenvalue 0. The eigenvector in eq. \ref{24} should be
identified with: 
\begin{equation}
\left ( 
\begin{array}{c}
\sin \psi \\ 
\cos \psi \cos \theta \\ 
\cos \psi \sin \theta
\end{array}
\right ),  \label{25}
\end{equation}
which, for $\theta=\frac {\pi}{4}$, has almost equal second and third
components, implying $U_{23} \simeq U_{33}$, in disagreement with our
assumption that the mixing angles of Dirac neutrinos with the corresponding
leptons are small.

Let us consider also the term proportional to $\mu_2 \,\, \mu_3$, and, for
brevity, define: 
\begin{eqnarray}
M & = & \tilde{N}_{33} \, \mu^2_3  \nonumber \\
\tilde{M} & = & \tilde{N}_{23} \, \mu_2 \mu_3.  \label{26}
\end{eqnarray}
In such a case, the neutrino mass matrix would be: 
\begin{eqnarray}
& M &\left ( \begin{matrix} {(U_L)^2_{13} & (U_L)_{13} (U_L)_{23}&
(U_L)_{13} (U_L)_{33} &\cr (U_L)_{13} (U_L)_{23}&(U_L)_{23}^2&(U_L)_{23}
(U_L)_{33} \cr (U_L)_{13} (U_L)_{33}& (U_L)_{23} (U_L)_{33}&(U_L)_{33}^2}
\end{matrix} \right ) \nonumber \\
 \\  \label{27} 
+ &\tilde{M}& \left ( \begin{matrix}{2(U_L)_{13} (U_L)_{12} & (U_L)_{13}
(U_L)_{22}+(U_L)_{12} (U_L)_{23} &(U_L)_{12} (U_L)_{33}+(U_L)_{13}(U_L)_{32}
\cr (U_L)_{13} (U_L)_{22}+(U_L)_{12} (U_L)_{23}& 2 (U_L)_{22}(U_L)_{23} &
(U_L)_{22}(U_L)_{33}+(U_L)_{23}(U_L)_{32} \cr
(U_L)_{12}(U_L)_{33}+(U_L)_{13}(U_L)_{32} &
(U_L)_{22}(U_L)_{33}+(U_L)_{23}(U_L)_{32} & 2(U_L)_{32} (U_L)_{33}}
\end{matrix} \right ) \nonumber 
\end{eqnarray}
which has a vanishing eigenvalue and the other two given by: 
\begin{equation}
\frac{M \pm \sqrt{M^2 + 4 \tilde{M}^2}}{2}  \label{28}
\end{equation}
By taking $M$ positive, one should identify 
\begin{equation}
\Delta m^2_a = \frac{1}{4} \; [ M + \sqrt{M^2 + 4 \tilde{M}^2} ]^2,
\label{29}
\end{equation}
while for $\Delta m^2_s$ ($<<$ $\Delta m^2_a$ ) one should have 
\begin{equation}
\Delta m^2_s \,= \, \frac{1}{4} [ \sqrt{M^2 + 4 \tilde{M}^2 } - M ]^2
\label{30}
\end{equation}

\begin{eqnarray}
\frac{\Delta m^2_s}{\Delta m^2_{a}} \simeq \frac{\Delta m^2_s \Delta m^2_{a}%
}{(\Delta m^2_{a})^2 + (\Delta m^2_s)^2} =  \nonumber \\
\frac{2 \tilde{M}^4}{2M^4 + 8 \tilde{M}^2 M^2 + 4 \tilde{M}^4} \simeq \left( 
\frac{\tilde{M}}{M } \right )^4  \label{31}
\end{eqnarray}
In the most favourable case, large angle MSW, it would give: 
\begin{equation}
\left ( \frac{\tilde{M}}{M} \right ) =\left( \frac{1.6 \cdot 10^{-5}}{3.5
\cdot 10^{-3}} \right )^{\frac{1}{4}} \sim \frac{1}{3.8}  \label{32}
\end{equation}

Let us see how large the non-diagonal matrix elements of $U_L$ should be in
order to give rise to a neutrino mass matrix, which has the property of
having almost equal matrix elements in $22$ and $33$ positions.We should
have 
\begin{equation}
M \, (U_L)^2_{23} + 2 \tilde{M} \, (U_L)_{23} (U_L)_{22} = M (U_L)^2_{33} +
2 \tilde{M} (U_L)_{33} (U_L)_{32}  \label{33}
\end{equation}
which is impossible to obtain with small values of the non-diagonal matrix
elements, since 
\begin{equation}
(U_L)_{23} (M (U_L)_{23} + 2 \tilde{M} (U_L)_{22} ) << (U_L)_{33} (M
(U_L)_{33} + 2\tilde{M} (U_L)_{32} )  \label{34}
\end{equation}
So the expansion with only the two terms is not able to reproduce a neutrino
mass matrix consistent with the experimental information. Therefore, we must
now consider further terms, i.e., those proportional to $\mu^2_2$ and to $%
\mu_1 \mu_3$. If we define: 
\begin{eqnarray}
\sigma & = & \tilde{N}_{22} \, \mu^2_2  \nonumber \\
\tau & = & \tilde{N}_{13}\, \mu_1 \, \mu_3  \label{35}
\end{eqnarray}
in the limit $(U_L)_{ij}= \delta_{ij}$, we would just find the matrix
proposed by Stech\cite{St} 
\begin{equation}
\left ( 
\begin{array}{ccc}
0 & 0 & \tau \\ 
0 & \sigma & \mu \\ 
\tau & \mu & M
\end{array}
\right )  \label{36}
\end{equation}

which, to reproduce the maximal mixing for $\nu_\mu$ and $\nu_\tau$,
requires the equality of the terms proportional to $\mu_2^2$ and $\mu_3^2$,
respectively.\newline
From this discussion it follows that the term proportional to $\mu_3^2$ does
not play a dominant role in the effective $\nu_L$ Majorana mass matrix not
only with respect to the one proportional to $\mu_2 \, \mu_3$, but also with
respect to the one proportional to $\mu_2^2$.For a diagonal lepton Dirac
matrix the term $\mu_3^2 $ appears in the expression of $M_{33}^R$,
multiplied by$({M^{L}}^{-1})_{33}$. So to reduce the contribution of $%
\mu_3^2 $ we take a small value for $({M^{L}}^{-1})_{33}$.\newline
In the following section we shall impose the vanishing of $({M^{L}}%
^{-1})_{33}$ and consequently of $(M^R)_{33}$ with a diagonal neutrino Dirac
matrix given by: 
\begin{equation}
m_\nu^D = \frac {m_\tau}{m_b} diag(m_u, m_c, m_t)  \label{37}
\end{equation}
\newline

\section{See-Saw model in SO(10) with diagonal neutrino Dirac matrix and $%
M_{33}^R = 0$}

\setcounter{equation}{0}

We have seen in the previous section that $M^L$, derived from a Dirac
neutrino mass matrix with the properties of the quark mass matrix,
hierarchical relationship for its eigenvalues and small mixing angles, is
not dominated by the term proportional to $(m_{\nu_\tau}^D)^2$,which is as
important as the term proportional to $(m_{\nu_\mu}^D)^2$.A way to implement
this property, the non-dominance of the term proportional to $%
(m_{\nu_\tau}^D)^2$, is found by considering the inverse of the first of
eqs. \ref{17}: 
\begin{equation}
M^R = - m_{\nu} ^D (M^L)^{-1} m_{\nu}^D.  \label{38}
\end{equation}
with $m_{\nu}^D$ given by eq. \ref{37}: 
\begin{eqnarray}
(\frac{m_b}{m_\tau m_u})^2 (M^R)_{11} & = & - \frac{1}{m_1} \, \cos^2 \psi
\cos^2 \chi - \frac{1 }{m_2} \, \cos^2 \psi \sin^2 \chi - \frac{1}{m_3} \,
\sin^2 \psi  \nonumber \\
\frac{(\frac{m_b}{m_\tau })^2} {m_u m_c} (M^R)_{12} & = & \frac{1}{\sqrt{2}%
m_1} \, \cos \psi \cos \chi (\cos \chi \, \sin \psi + \sin \chi )\nonumber \\
& & - \frac{1}{%
\sqrt{2}m_2} \,\cos \psi \sin \chi (\cos \chi - \sin \psi \, \sin \chi ) 
 - \frac{1}{2\sqrt{2}m_3} \, \sin(2\psi)  \nonumber \\
\frac{(\frac{m_b}{m_\tau })^2} {m_u m_t} (M^R)_{13} & = & \frac{1}{\sqrt{2}%
m_1} \, \cos \psi \cos \chi (\cos \chi \, \sin \psi - \sin \chi ) \nonumber \\
& & + \frac{1}{%
\sqrt{2}m_2} \, \cos \psi \sin \chi (\cos \chi + \sin \psi \, \sin \chi ) 
 -\frac{1}{2 \sqrt {2}m_3} \,\sin(2\psi)  \nonumber \\
(\frac{m_b}{m_\tau m_c})^2 (M^R)_{22} & = & - \frac{1}{2m_1} (\cos \chi \sin
\psi + \sin \chi )^2  \\ \label{39}
& & - \frac{1}{2m_2} \, (\cos \chi - \sin \psi \, \sin \chi
)^2 - \frac{1}{2m_3} \, \cos^2 \psi  \nonumber \\
\frac{(\frac{m_b}{m_\tau })^2} {m_c m_t}(M^R)_{23} & = &   \frac{1}{2m_1} 
(\sin^2 \chi - \cos^2 \chi \sin^2 \psi ) \nonumber \\
& & + \frac{1}{2m_2} \, (\cos^2 \chi -
\sin^2 \psi \, \sin^2 \chi ) - \frac{1}{2m_3} \, \cos^2 \psi  \nonumber \\
(\frac{m_b}{m_\tau m_t})^2 (M^R)_{33} & = & - \frac{1}{2m_1} (\sin \chi -
\cos \chi \sin \psi )^2 \nonumber \\
& & - \frac{1}{2m_2} \, (\cos \chi + \sin \chi \, \sin
\psi )^2 - \frac{1}{2m_3} \, \cos^2 \psi  \nonumber
\end{eqnarray}

From eqs. \ref{39}, it is easy to see that 
\begin{equation}
|(M^{R})_{11}|<\frac{(\frac{m_{\tau }m_{u}}{m_{b}})^{2}}{|m_{1}|}  \label{40}
\end{equation}
which implies \cite{fk}, in the absence of cancellations for $(M^{R})_{33}$ 
\footnote{{Since the ratios of the quark masses have a negligeable
dependence on the scale, we can take the values given in \cite{fk}}}: 
\begin{equation}
|\frac{(M^{R})_{11}}{(M^{R})_{33}}|\simeq 2\frac{m_{u}^{2}}{m_{t}^{2}}\simeq
2\left( \frac{\frac{8}{3}{\rm MeV}}{180{\rm GeV}}\right) ^{2}\simeq 4\cdot
10^{-10}  \label{41}
\end{equation}
To prevent the appearence of such unnatural small factor we impose the
vanishing of the r.h.s of last equation in \ref{39}: 
\begin{equation}
\frac{(\sin \chi -\cos \chi \sin \psi )^{2}}{m_{1}}+\frac{(\cos \chi +\sin
\psi \sin \chi )^{2}}{m_{2}}+\frac{\cos ^{2}\psi }{m_{3}}=0  \label{42}
\end{equation}
which requires that not all the $m_{i}$'s have the same sign and 
\begin{eqnarray}
\frac{m_{b}^{2}}{m_{\tau }^{2}m_{c}m_{t}}(M^{R})_{23} & = & \frac{\sin 2\chi \sin
\psi }{2}(\frac{1}{m_{1}}-\frac{1}{m_{2}})\nonumber \\
& & -(\sin \psi )^{2}(\frac{(\cos \chi
)^{2}}{m_{1}}+\frac{(\sin \chi )^{2}}{m_{2}})-\frac{(\cos \psi )^{2}}{m_{3}}
\label{43}
\end{eqnarray}
By neglecting the terms proportional to powers of $\sin \psi $, we get 
\begin{equation}
|(M^{R})_{23}|\leq 7\cdot 10^{11}{\rm GeV}  \label{44}
\end{equation}
as in \cite{ABFRT}.

With $\sin \psi=0$, eqs. \ref{39} read: 
\begin{eqnarray}
(\frac{m_b}{m_\tau m_u})^2 (M^R)_{11} & = & - \frac{1}{m_1} \cos^2 \chi 
\nonumber \\
\frac{(\frac{m_b}{m_\tau })^2} {m_u m_c} (M^R)_{12} & = & \frac{\cos \chi
\sin \chi }{\sqrt{2}}(\frac{1}{m_1} - \frac{1}{m_2})  \nonumber \\
\frac{(\frac{m_b}{m_\tau })^2} {m_u m_t} (M^R)_{13} & = & \frac{\cos \chi 
\sin \chi }{\sqrt{2}}(\frac{1}{m_1} + \frac{1}{m_2}) \label{45}  \\
(\frac{m_b}{m_\tau m_c})^2 (M^R)_{22} & = & - \frac{1}{2m_1} \sin^2 \chi - 
\frac{1}{2m_2} \cos^2 \chi - \frac{1}{2m_3}  \nonumber \\
\frac{(\frac{m_b}{m_\tau })^2} {m_c m_t}(M^R)_{23} & = & \frac{1}{2m_1}
\sin^2 \chi + \frac{1}{2m_2} \cos^2 \chi - \frac{1}{2m_3}  \nonumber \\
(\frac{m_b}{m_\tau m_t})^2 (M^R)_{33} & = & - \frac{1}{2m_1} \sin^2 \chi - 
\frac{1}{2m_2} \cos^2 \chi - \frac{1}{2m_3} \nonumber 
\end{eqnarray}
In order to get $(M^R)_{33}=\,0$, one should have 
\begin{equation}
\frac{\sin^2 \chi}{m_1} + \frac{\cos^2 \chi}{m_2} + \frac{1}{m_3} = 0
\label{46}
\end{equation}

We look for its solutions with $m_i$ 's constrained by the relations:

\begin{eqnarray}
m_{2}^{2} &=&m_{1}^{2}+\Delta m_{s}^{2}  \nonumber \\
m_{3}^{2} &=&m_{1}^{2}+\Delta m_{a}^{2}  \label{47}
\end{eqnarray}
With this we do not want to impose that $(M^{R})_{33}$ vanishes exactly.
But, once a solution with no large differences between the matrix elements
of $M^{R}$ has been found, we can allow for a value of $(M^{R})_{33}$ of the
same order of the other matrix elements, with a small change for the
left-handed neutrino Majorana mass matrix.\newline
It is obvious that to obey eq. \ref{46} the $m_{i}$ 's cannot have the same
sign and, as $|m_{3}|>|m_{2}|>|m_{1}|$, $m_{1}$ and $m_{2}$ should have
opposite signs.This fact has been pointed out in \cite{F}. From eq. \ref{46}
we get: 
\begin{equation}
\tan ^{2}\chi =-\frac{m_{1}(m_{3}+m_{2})}{m_{2}(m_{3}+m_{1})}  \label{48}
\end{equation}
which,when $|m_{2}|<<|m_{3}|$ simplifies to 
\begin{equation}
\tan ^{2}\chi =-\frac{m_{1}}{m_{2}}  \label{49}
\end{equation}
in agreement with the expectation that large mixing angles correspond to
almost degenerate masses of the mixed states \cite{ELLN}. One has another
approximate solution with almost degenerate lightest neutrino mass
eigenstates ($|m_{2}|\simeq -m_{1}=m$) 
\begin{equation}
m=\cos (2\chi )m_{3}  \label{50}
\end{equation}
These approximate solutions correspond to exact solutions of eq. \ref{46}. In
fact, with equal signs for $m_{2}$ and $m_{3}$ we may write eq. \ref{46} in
the form 
\begin{equation}
\sin \chi ^{2}+\cos \chi ^{2}\frac{m_{1}}{m_{2}}+\frac{m_{1}}{m_{3}}=0
\label{51}
\end{equation}
which has one and only one solution for negative $m_{1}$ since its l.h.s. is
an increasing function of $m_{1}$ in the range $(-\infty ,0)$ varying from $%
-2\cos ^{2}\chi $ to $\sin ^{2}\chi $ \footnote{%
Notice that eqs. \ref{47} imply that when $\left| m_{1}\right| $ goes to $%
\infty $ the ratios $\left| \frac{m_{1}}{m_{2}}\right| $ and $\left| \frac{
m_{1}}{m_{3}}\right| $ go to 1.}. In the case of the same sign for $m_{1}$
and $m_{3}$ it is convenient to rewrite eq. \ref{46} in the form

\begin{equation}
\sin ^{2}\chi \frac{m_{2}}{m_{1}}+\cos ^{2}\chi +\frac{m_{2}}{m_{3}}=0
\label{52}
\end{equation}
With its l.h.s., in the range ($-\infty $, $-\sqrt{\Delta m_{s}^{2}})$ for $%
m_{2}$, going from $-2(\sin ^{2}\chi )$ to $-\infty $. But, if $\cos^{2}\chi $
is sufficiently larger than $\sin ^{2}\chi $, can reach positive values,
implying two solutions for eq. \ref{46}, approximately given by eqs. \ref{49}
and \ref{50}, respectively.\newline
For the MSW solutions one has to take $\cos ^{2}\chi >\sin^{2}\chi $ and
therefore one has three solutions. For the vacuum solutions, which allows
both signs for $\cos (2\chi )$, one has also a solution with $\sin^{2}\chi
>\cos ^{2}\chi $. We wish to extend the previous analysis to consider non
vanishing values of $|\sin \psi |$ . In that case, when 
\begin{equation}
(\sin \chi -\sin \psi \cos \chi )^{2}>(\cos \chi +\sin \psi \sin \chi )^{2}
\label{53}
\end{equation}
one has only one solution of eq. \ref{46}; instead, when the r.h.s. of the
inequality \ref{53} is sufficiently larger than the l.h.s. + 1, there are
two solutions. We have performed a numerical study of the solutions of eq. 
\ref{43}  with $|\sin \psi |=k (.075)$, k=0,...,3. To get a general view of
our solutions, we give in Table II, for each solar neutrino oscillation
scenario, the matrix where the smallest value for the ratios of the moduli
of non-vanishing matrix elements of $M^{R}$ (which is, with the only
exeption of VACL solution, $\frac{M_{11}^{R}}{M_{23}^{R}}\ $) takes the
largest value and the ones where the max (ie. highest) matrix element of $%
M^{R}$ takes the smallest or largest value. From Table II we see that the
matrices with smallest excursion for their matrix element correspond to
small values for ${m_{1}}^{2}$ and to $\psi =0$, with the only exception of
the LMW solution, where $|\sin \psi |=0.075$. The matrices with the smallest
value of $M_{23}^{R}$ are found for the large ${m_{1}}^{2}$ solutions.
Finally the largest values of $M_{23}^{R}$ are found at the boundary of the
allowed values for $\psi $ and in correspondence of the small ${m_{1}}^{2}$
solutions. Since our motivation for setting $M_{33}^{R}=0$ is also based on the demand
of dealing with $M^{R}$ matrices with not so disparate orders of magnitude for its
matrix elements, we
favour small $\psi $ and ${m_{1}}^{2}$ choices (in particular, the small
MSW solution with the ratio between the smallest
and the largest matrix elements of $M^{R}$ given by $4.3\cdot 10^{-4}$ ).
 We never find  neutrino masses large enough to be of
cosmological relevance. The maximum value found for $\sum_{i}\left|
m_{i}\right|$ being .6 eV, corresponding to the boundary of allowed values
of the small angle MSW solution.

\section{Conclusions}

The hypothesis of a vanishing $M_{33}^{R}$, assumed in\cite{js}\cite{St}\cite
{ABFRT} and here motivated by the requirement of not having too large fine
tunings in the see-saw formula may be implemented in all five solutions to
the ''solar neutrino problem''.In general one has solutions with large $\sim
(\Delta m_{a}^{2}$) or small $\sim (\Delta m_{s}^{2})$ for the square mass
of the lightest neutrino mass eigenstate. For large values of $m_{1}^{2}$
the max matrix element of $M^{R}$ ($M_{23}^{R}$) has a value $\sim 7\cdot
10^{11}$ GeV, a scale found in a recent work \cite{ABFRT} and in agreement
with the spontaneous scale of B-L symmetry breaking found in a SO(10) model 
\cite{BP} with SU(4) x SU(2) x SU(2) intermediate symmetry \cite{ps}.
Moderate values for the matrix elements of $M^{R}$ have been also found in 
\cite{js} \cite{St} and recently by Akhmedov, Branco and Rebelo \cite{ABR}.
For small values for $m_{1}^{2}$ one still finds the same order of magnitude
for the large angle MSW solution in the entire range of the allowed values
for $\psi $; the same happens for the other MSW solutions for $\psi $ = 0.
The other MSW solutions correspond to larger values for the scales,
especially for the largest allowed values of $|\sin \psi |$. The solutions
corresponding to small $m_{1}^{2}$ at $\psi $=0 have the appealing feature
of having not too different orders of magnitude for the matrix elements of $%
M^{R}$.

{\bf Acknowledgements:} One of us (F.B.) acknowledges G. Fiorentini for
interesting discussions.

\newpage

~~~~
\includegraphics{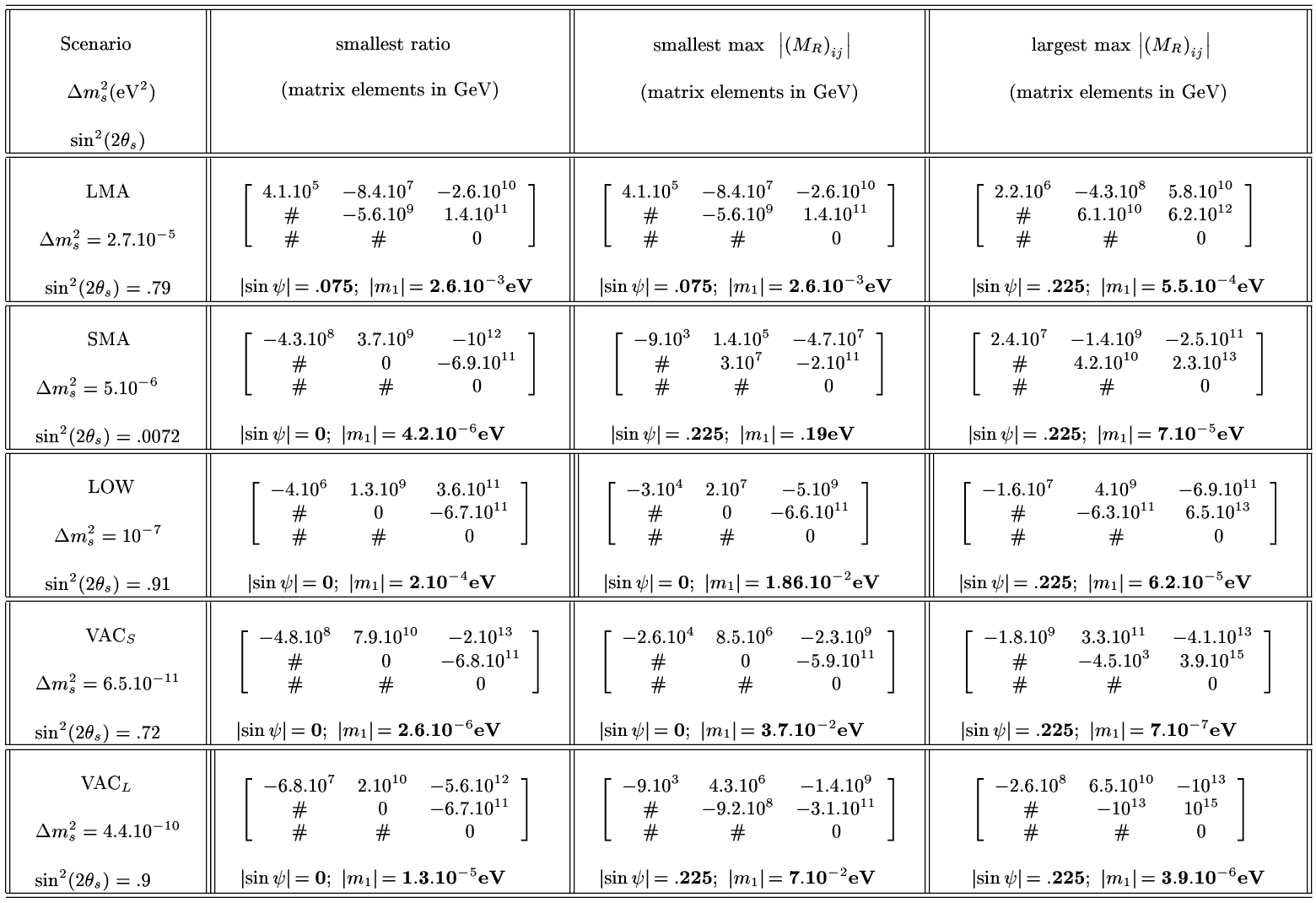}

\small
\vskip 19.5 truecm
\centerline{Table II}
\vskip 0.4 truecm

\noindent The matrices $M_{R}$  obeying eq. 3.6 for the five scenarios.
For each solution the corresponding values of $\left| \sin \psi
\right| $ and $\left| m_{1}\right| $ are given.
As  $M_{R}$ is symmetric, repeated  matrix elements are denoted by \#.

\end{document}